# Reduced Optical Gain Threshold by Carrier Multiplication in Semiconductor Perovskite Nanocrystals

Zhen Zhang[1†], Encheng Sun[1†], Jian Li[1†], Chunfeng Zhang[1], Fengrui Hu[2*], Min Xiao[1], and Xiaoyong Wang[1*]

[1]*National Laboratory of Solid State Microstructures, School of Physics, and Collaborative Innovation Center of Advanced Microstructures, Nanjing University, Nanjing 210093, China*

[2]*College of Engineering and Applied Sciences, and MOE Key Laboratory of Intelligent Optical Sensing and Manipulation, Nanjing University, Nanjing 210093, China*

[*]Correspondence to F.H. (frhu@nju.edu.cn) or X.W. (wxiaoyong@nju.edu.cn)

[†]These authors contributed equally to this work

**Carrier multiplication (CM) describes a strong charge-carrier interaction process in semiconductor colloidal nanocrystals (NCs), wherein two band-edge excitons are simultaneously created by an absorbed photon with at least twice the bandgap energy (2 $E_g$). While being fundamentally intriguing, it has been exclusively utilized to enhance the light-to-electricity conversion efficiencies in the photodetector and solar-cell devices. In this report, we have synthesized the core/shell perovskite $FAPbI_3/NdF_3$ NCs with a biexciton recombination lifetime of ~3.9 ns, and demonstrated that a CM efficiency of ~25.7% can be achieved under the ~355 nm laser excitation (~2.21 $E_g$). This CM occurrence leads to a two-fold reduction in the optical gain threshold, as compared to that obtained under the ~640 nm laser excitation (~1.23 $E_g$). When combined with the**

**single-exciton and zero-threshold optical gain schemes previously developed for semiconductor colloidal NCs, the CM effect introduced here would further mitigate the optical-pumping requirement for the routine operation of continuous-wave lasing.**

Semiconductor colloidal nanocrystals (NCs) have been attracting great research attentions as one of the leading materials for laser application, owing to their low cost synthesis, flexible solution processibility, high fluorescence efficiency, tunable bandgap energy and narrow gain spectrum[1,2]. Ever since the first report of amplified spontaneous emission (ASE) from semiconductor colloidal NCs[3], they have been optically pumped to realize both solid- and liquid-phase lasing operations inside various optical cavities such as with the distributed feedback[4], microsphere[5,6], vertical surface-emitting[7] and Littrow[8,9] structures. For the eventual goal of large-scale commercialization, it is imperative to routinely realize colloidal NC lasing under the continuous-wave (CW) pumping condition, which requires the reduction (elongation) of optical gain threshold (lifetime) during the population inversion process of photogenerated excitons. Associated with a non-unity degeneracy of the band-edge state, the colloidal NCs have to be at least moderately pumped to prepare biexcitons for the population inversion purpose; meanwhile, the biexciton would transfer energy to the single exciton in a nonradiative Auger process before its radiative recombination[10]. For the traditional metal-chalcogenide[11] and the emerging perovskite[12] colloidal NCs, the biexciton Auger process normally occurs at several to hundreds of picoseconds, thus necessitating the employment of ultrashort laser pulses to promote ASE within this limited time window. The situation becomes ever worse for the electrically-pumped lasing operation, since the optical gain parameters need to be further improved due to the unavoidable introduction of optically-lossy charge conducting layers[13].

To elongate the optical gain lifetime, a feasible strategy is to adopt a heterogeneous structure with the type-II energy-level alignment, which has been achieved in the core/shell CdSe/CdS[14] and the core/crown CdSe/CdTe[15] colloidal NCs. Therein, the excited-state electron and hole are separated at different spatial locations, with their wavefunction overlap being significantly reduced to benefit the suppression of biexciton Auger recombination. In fact, the requirement for population inversion can also be mitigated in this kind of type-II core/shell NC structures, wherein the repulsive exciton-exciton interaction can be enhanced to yield single-exciton optical gain by reducing spectral overlap of the absorption and emission states[7,16,17]. Alternatively, the colloidal NCs can be chemically doped and then optically excited to favor the dominant formation of charged excitons, which are capable of bringing about a zero-threshold optical gain when the ground-state absorption is completely blocked[18,19]. Despite the above and some other research efforts, such as lifting the valence band degeneracy by means of uniform biaxial strain, the optically-pumped CW lasing is still rarely reported for the colloidal NCs[20]. Furthermore, although the electrically-driven ASE has been recently achieved in the colloidal NCs[21], its further manipulation into lasing operation is still a challenging task yet to be fulfilled by the research community. Since the biexciton Auger lifetime can now be extended to a high value beyond several nanoseconds[9,14], the routine optically-pumped CW or first electrically-driven lasing is critically dependent on a further reduction of the optical gain threshold in the colloidal NCs[17].

In this report, we have synthesized the core/shell perovskite FAPbI$_3$/NdF$_3$ NCs with a type-II energy-level alignment, with the resultant biexciton Auger lifetime of ~3.9 ns being long enough for the persistence of optical gain in the optically-pumped ASE process. Meanwhile, this biexciton Auger lifetime is still relatively short to imply the existence of a moderate charge-carrier interaction to support the CM occurrence. These FAPbI$_3$/NdF$_3$ NCs demonstrate a CM efficiency of ~25.7% upon the ~355 nm laser excitation with a photon energy ~2.21 times that of the bandgap transition (~2.21 $E_g$). With <$N$> being the average number of photons absorbed per laser pulse by a single FAPbI$_3$/NdF$_3$ NC, this high CM efficiency is converted to a low optical gain threshold of <$N$> = ~0.68, as compared to that of <$N$> = ~1.20 obtained from the ~640 nm laser excitation with a photon energy of ~1.23 $E_g$. As a consequence, the ASE threshold is reduced from <$N$> = ~1.35 to ~0.85 when the laser excitation wavelength is switched from ~640 to ~355 nm, thus signifying the emergence of a potent strategy of utilizing CM to advance the current lasing studies of semiconductor colloidal NCs.

When assuming a two-fold energy degeneracy in the band-edge state, we first plot schematically in Fig. 1a how a single colloidal NC interacts with the incident laser with a photon energy of $hv < 2 E_g$. At the low laser power, an electron is promoted from the valence band to a lower-lying excited state in the conduction band, and then relaxes to the band-edge state to finish the optical absorption process (i). This single NC is optically transparent for the photons emitted from the neighbouring NCs with the energy of $hv = E_g$ (ii), since the probability for another electron being excited to the band-edge state is equal to that for the existing one to undergo stimulated emission (SE). Only when a high laser power is employed to prepare two electrons in the band-edge state can the optical gain feature be rendered to the single NC (iii), from which two emitted photons can be induced by one incident photon all with the same energy of $h\upsilon = E_g$. After describing the electronic behaviors of a single NC excited strongly at $h\upsilon < 2 E_g$, we propose in Fig. 1b that the optical gain can be alternatively prepared by a low-power laser whose photon energy of $h\upsilon > 2 E_g$ is sufficient to trigger the CM effect. In this case, a valence-band electron is promoted to a higher-lying excited state in the conduction band upon photon absorption (iv). After proceeding a little bit in the downward relaxation, this electron can arrive directly at the band-edge state, to which another valence-band electron is simultaneously excited after capturing the released energy (v). The band-edge state is now occupied by two electrons to realize optical gain after a single-photon absorption event (vi), thanks to the CM effect as a result of enhanced charge-carrier interaction in the quantum-confined colloidal NC[22].

To investigate whether the CM mechanism proposed above is feasible for the reduction of optical gain threshold, we focus here on the perovskite core/shell $FAPbI_3/NdF_3$ NCs synthesized according to a standard procedure reported very recently[23] (see Methods). As demonstrated in Supplementary Fig. 1, these $FAPbI_3/NdF_3$ NCs possess an average edge length of ~18.5 nm with the band-edge absorption and emission peaks located at ~1.62 eV (~765.4 nm) and ~1.58 eV (~784.6 nm), respectively. For the single-particle optical studies at room temperature (see Methods), one drop of the diluted sample solution is spin-coated on top of a $SiO_2$ substrate to yield the well-isolated single $FAPbI_3/NdF_3$ NCs, which are excited by the picosecond pulsed laser with the output wavelength of ~355, ~366, ~385 or ~640 nm. These four laser wavelengths correspond to the photon energies of ~2.21, ~2.14, ~2.04 and ~1.23 $E_g$, respectively, as compared to that of ~784.6 nm measured for the band-edge emission peak. The laser excitation power can be varied to yield a specific value for the aforementioned <$N$> (see Supplementary Fig. 2), which is also the average number of excitons created per pulse in a single FAPbI$_3$/NdF$_3$ NC.

For a representative single FAPbI$_3$/NdF$_3$ NC excited at ~640 nm with <$N$> = ~0.1, the PL intensity measured in Fig. 2a shows a time-dependent flickering behavior with a single-peak Gaussian distribution. This signifies a slow transition between the neutral and charged excitons in the single NC, the latter of which is associated with a high fluorescent efficiency due to the suppressed Auger recombination[23]. When the ~640 nm laser power is further increased in Fig. 2b to set <$N$> = ~0.5, the flickering behaviour becomes more drastic due to the increased probability of biexciton Auge ionization to create more charged excitons[24], leading to a slight broadening of the single Gaussian peak for the PL intensity distribution. In Fig. 2c, we provide the PL intensity time trace measured for the same single NC under the ~355 nm laser excitation with <$N$> = ~0.1, which is quite similar to that shown in Fig. 2b in terms of the drastic flickering behaviour and the broadened Gaussian distribution.

After obtaining the above three PL intensity time traces, we next construct the corresponding PL decay curves in Fig. 2d using those photons emitted from the neutral excitons as marked by the grey "on" boxes in Fig. 2a-2c. As a consequence, the possible interference of charged excitons to the estimation of CM efficiency[25,26] can be effectively removed[27]. Under the ~640 nm laser excitation with <$N$> = ~0.1 to create mainly single excitons, the PL decay curve can be well fitted with a single-exponential lifetime of ~47.1 ns ($\tau_x$) from their radiative recombination. When <$N$> is increased to ~0.5 to create both single excitons and biexcitons, the PL decay curve has to be fitted by a biexponential function of $A_x e^{-t/\tau_x} + A_{xx} e^{-t/\tau_{xx}}$, with $A_x$ ($A_{xx}$) and $\tau_x$ ($\tau_{xx}$) being the amplitude and the value of the slow (fast) lifetime component, respectively. In addition to the slow lifetime of $\tau_x$ = ~46.4 ns from the radiative recombination of single excitons, the fast lifetime of $\tau_{xx}$ = ~3.3 ns, with an amplitude ratio of $A_{xx}/(A_x + A_{xx})$ = ~26.2%, should be dominated by the nonradiative Auger recombination of biexcitons. Based on the same PL decay measurements on 45 single FAPbI$_3$/NdF$_3$ NCs, the biexciton lifetimes of $\tau_{xx}$ are averaged at ~3.9 ns to imply a moderate suppression of nonradiative Auger recombination in the core/shell structure.

Still in Fig. 2d for the ~355 nm laser excitation with <$N$> = ~0.1, this single FAPbI$_3$/NdF$_3$ NC has already shown a biexponential PL decay with the slow and fast lifetimes of $\tau_x$ = ~46.8 ns and $\tau_{xx}$ = ~3.6 ns, respectively. The latter fast lifetime is comparable to that of ~3.3 ns measured under the ~640 nm laser excitation with <$N$> = ~0.5, albeit with a larger amplitude ratio of ~43.5% to imply a high biexciton generation efficiency in the CM process. Of special note is that when the laser excitation wavelength is switched back to ~640 nm with <$N$> = ~0.1, a single-exponential PL decay curve is recovered with the fitted lifetime of ~45.7 ns (Supplementary Fig. 3). This lifetime is similar to the original one of ~47.1 ns measured under the same excitation condition, suggesting that no degradation occurs for the single FAPbI$_3$/NdF$_3$ NC under the ~355 nm laser excitation with a photon energy of ~2.21 $E_g$.

According to the formular of $\beta = \frac{A_{xx}(\tau_x - \tau_{xx})}{A_x(3\tau_x - 4\tau_{xx}) - A_{xx}\tau_{xx}}$ from a theoretical model established previously[27], the CM efficiency ($\beta$) for biexciton generation is calculated to be ~27.3% for the specific single FAPbI$_3$/NdF$_3$ NC studied in Fig. 2a-2d. To reveal further information on the CM process in this single FAPbI$_3$/NdF$_3$ NC, we change the laser excitation wavelengths to ~366 nm (~2.14 $E_g$) and ~385 nm (~2.04 $E_g$) still with <*N*> = ~0.1 (see the PL intensity time traces and PL decay curves in Supplementary Fig. 4 and Fig. 5, respectively). For the ~366 nm laser excitation, the CM efficiency is estimated to be ~22.6% from a biexponential fitting of the PL decay curve, which is obviously smaller than that of ~27.3% under the ~355 nm laser excitation due to the decreased excitation photon energy[28]. As to the ~385 nm laser excitation, the obtained PL decay curve is well fitted with a single-exponential lifetime of ~46.2 ns from the radiative recombination of single excitons, since the photon energy of ~2.04 $E_g$ is not large enough to tigger the CM effect for the generation of biexcitons. Based on the statistical PL measurements on 52 single FAPbI$_3$/NdF$_3$ NCs under the ~355 nm (~2.21 $E_g$) laser excitation, the CM efficiencies plotted in the histogram of Fig. 2e are averaged at ~25.7%. This efficiency is significantly larger than that of ~8.4% obtained previously for the single CdSe/ZnS NCs excited at ~2.22 $E_g$[27], which can be attributed to the long relaxation time of hot charge carriers in the perovskite materials to facilitate the CM occurence[29,30]. In Fig. 2f, we additionally provide a statistical histogram for the distribution of CM efficiencies estimated under the ~366 nm (~2.14 $E_g$) laser excitation, which are averaged at ~18.1% for the 37 single FAPbI$_3$/NdF$_3$ NCs studied in our experiment.

After confirming that the CM effect can really be triggered in the single FAPbI$_3$/NdF$_3$ NCs under the ~355 nm laser excitation, we move forward to investigate whether it can be utilized to reduce the optical gain threshold. To this end, one drop of the concentrated sample solution is spin-coated onto a $SiO_2$ substrate to form a solid film of the ensemble $FAPbI_3/NdF_3$ NCs. For the transient absorption (TA) measurement (see Methods), the ensemble $FAPbI_3/NdF_3$ NCs are excited by the ~355 or ~640 nm output beam from an optical oscillation amplifier (OPA), which is operated at the repetition frequency of ~1 KHz with a pulse width of ~100 fs. In Fig. 3a, we plot a 2D image to show time-dependent evolution of the TA spectra from ~200 fs to ~900 ps, which are obtained for the $FAPbI_3/NdF_3$ NCs under the ~355 nm laser excitation with $<N>$ = ~10.32. Therein, the ground-state bleaching (GSB) peak has the same energy of ~1.62 eV as that measured for the band-edge absorption peak (Supplementary Fig. S1b), and it is accompanied by a higher-energy tail with a fast-decaying trend over the time delay, corresponding to the band filling and thermal relaxation processes of photogenerated hot charge carriers[31]. On the even higher-energy side of GSB, there exists a broad photo-induced absorption (PIA) signal (PIA1, ~1.85-2.25 eV) due to the electronic transition of charge carriers from the band-edge to the exited states[32].

On the lower-energy side of GSB in the TA spectra of Fig. 3a, another PIA peak denoted by PIA2 can be additionally resolved at the energy of ~1.59 eV, which is slightly higher than that of ~1.58 eV measured for the band-edge emission peak from single excitons (Supplementary Fig. S1b). This observation suggests the underlying origin of PIA2 from biexcitons with a negative binding energy, which is caused by enhanced exciton-exciton repulsion in the core/shell $FAPbI_3/NdF_3$ NCs with a type-II energy-level alignment[23]. It can be further seen from Fig. 3a that an SE peak is converted from PIA2 and then arrives at the highest amplitude after the time delays of ~1 and ~4 ps, respectively (Supplementary Fig. 6), which are comparable to those time scales commonly reported in the biexciton optical gain studies of semiconductor colloidal NCs[3,33]. Besides $<N>$ = ~10.32 employed in Fig. 3a, we have also measured time-dependent evolutions of the TA spectra at other $<N>$ values. For each $<N>$ at the time delay of ~4 ps, a TA spectrum ($\Delta\alpha$) is extracted and combined with the steady-state absorption spectrum ($\alpha_0$) to yield the nonlinear absorption spectrum of $\Delta\alpha + \alpha_0$, which would adopt a negative value after the optical gain has been generated[8,9,18]. These nonlinear absorption spectra are plotted in Fig. 3b with the increasing $<N>$ values from ~0.43 to ~10.32, showing the gradual appearance of optical gain to eventually cover the energy range from ~1.58-1.63 eV.

For the comparison purpose, we plot a 2D image in Fig. 3c to show time-dependent evolution of the TA spectra measured for the $FAPbI_3/NdF_3$ NCs under the ~640 nm laser excitation with $<N>$ = ~9.84, wherein the same GSB, PIA1 and PIA2 features as those in Fig. 3a can be observed. For this ~640 nm laser excitation at the time delay of ~4 ps, we plot in Fig. 3d the nonlinear absorption spectra of $\Delta\alpha + \alpha_0$ at the increasing $<N>$ values, showing that the optical gain is associated with a relatively narrower energy range of ~1.58-1.61 eV than that obtained in Fig. 3b for the ~355 nm laser excitation. Based on the two sets of nonlinear absorption spectra plotted in Fig. 3b and 3d, we extract the $<N>$-dependent $\Delta\alpha + \alpha_0$ values at the biexciton peak of ~1.59 eV and plot them in Fig. 3e. They approach zero values at $<N>$ = ~1.20 and ~0.68 for the ~640 and ~355 nm laser excitations, respectively, thus confirming that the optical gain threshold can indeed be significantly reduced by the CM effect. By extracting from Fig. 3a and 3c the TA time traces ($\Delta\alpha$) at the biexciton peak of ~1.59 eV and dividing them by $-\alpha_0$, we obtain the corresponding curves in Fig. 3f with $-\Delta\alpha/\alpha_0 > 1$ signifying the maintenance of optical gain[3,14,19]. Compared to that of ~440 ps estimated under the ~640 nm laser excitation with $<N>$ = ~9.84, the optical gain lifetime is favorably elongated to ~732 ps under the ~355 nm laser excitation with $<N>$ = ~10.32.

Now equipped with a thorough understanding of optical gain in the ensemble $FAPbI_3/NdF_3$ NCs in a solid film, we move forward to explore their ASE properties using an edge collection geometry at room temperature (see Methods). As shown in Fig. 4a for the $FAPbI_3/NdF_3$ NCs exited at ~640 nm, the PL peak measured at $<N>$ = ~0.15 is located at ~1.58 eV with a full width at half maximum (FWHM) of ~97.9 meV, and there appears a sharp ASE peak between ~1.58-1.59 eV after the further increase of $<N>$ to higher values. The spectral integrated intensities and peak FWHMs are plotted in Fig. 4b, from which the ASE threshold is estimated to be $<N>$ = ~1.35 at the cross point where they transit to the superlinear increase and the significant narrowing, respectively. When the same $FAPbI_3/NdF_3$ NCs are excited at ~355 nm in Fig. 4c, the ASE peak shows a much quicker emergence with the increase of $<N>$ and is located at a slightly higher energy position of ~1.59 eV, as compared to that measured in Fig. 4a for the ~640 nm laser excitation. According to the spectral integrated intensities and peak FWHMs plotted in Fig. 4d, the ASE threshold is now reduced to a lower value of $<N>$ = ~0.85 due to the occurrence of CM in the $FAPbI_3/NdF_3$ NCs excited at a photon energy of ~2.21 $E_g$. The ASE thresholds of $<N>$ = ~1.35 and ~0.85 measured at ~640 and ~355 nm are relatively higher than the corresponding optical gain thresholds of $<N>$ = ~1.20 and ~0.68, and their further reductions can be achieved by improving the optical quality of the solid film and increasing the volume fraction of the contained $FAPbI_3/NdF_3$ NCs.

In summary, we have synthesized the core/shell perovskite $FAPbI_3/NdF_3$ NCs with a biexciton lifetime of ~3.9 ns, and demonstrated that a CM efficiency of ~25.7% can be achieved under the ~355 nm laser excitation with a photon energy of ~2.21 $E_g$. This efficient CM process leads to the significantly-reduced optical gain and ASE thresholds of $<N>$ =~0.68 and ~0.85, respectively, as compared to those of $<N>$ =~1.20 and ~1.35 measured under the ~640 nm laser excitation with a photon energy of ~1.23 $E_g$. Given the fact that the CM effect in semiconductor colloidal NCs is now being exclusively studied towards the enhancement of power conversion efficiencies in the photodetector and solar-cell devices[34-37], our current work has rendered it a novel research direction to mitigate the lasing requirement by the creation of optical gain after the absorption of only a single near-UV photon. The CM effect is compatible with the single-exciton[17] or zero-threshold[18] mechanism already developed in the literature for semiconductor colloidal NCs, and they can be synergistically manipulated to achieve an optical gain threshold that is low enough to support the routine operation of optically-pumped CW lasing. It would be even more exciting to trigger the CM effect by electrically injecting high-energy (>2 $E_g$) charge carriers into the optical gain material of semiconductor colloidal NCs, with the as-lowered pumping threshold being beneficial to provide a potent driving force for the realization of laser diode operation.

**Methods**

**Sample synthesis.** To get the FA-oleate precursor, 0.064 g formamidinium acetate was added to 2 mL oleic acid, with the resulting solution being heated at the temperature of 80 °C for 20 min. For the PbI$_2$-ligand precursor, 0.046 g PbI$_2$, 100 μL oleic acid and 150 μL oleylamine were added to 10 mL toluene, with the resulting solution being stirred at the temperature of 100 °C for 1 h. As to the NdF$_3$-ligand precursor, 0.04 g of the neodymium fluoride powder was added to 2 mL isopropanol, with the resulting solution being stirred for 2 h at room temperature. Thereafter, 90 μL of the FA-oleate precursor and 1 mL of the PbI$_2$-ligand precursor were mixed together, with the resulting solution being stirred vigorously for 3 min and then centrifuged at 9000 rpm for 5 min. The obtained precipitate was redispersed in 1 mL hexane and centrifuged at 3000 rpm for 3 min to get the supernatant containing the core-only FAPbI$_3$ NCs. Finally, 20 μL of the NdF$_3$-ligand precursor was injected into the above solution of the FAPbI$_3$ NCs, yielding the core/shell FAPbI$_3$/NdF$_3$ NCs after 1 h of natural reaction.

**Single-particle measurement.** One drop of the diluted sample solution was spin-coated onto a SiO$_2$ substrate to form a solid film with the well-isolated single FAPbI$_3$/NdF$_3$ NCs, which were excited at ~640 nm by a picosecond pulsed laser operated at the repetition frequency of ~5 MHz. These single FAPbI$_3$/NdF$_3$ NCs could also be excited at ~355, ~366 or ~385 nm by the SHG (second harmonic generation) output of a picosecond Ti:Sapphire laser, whose repetition rate had been converted from ~76 to ~4.75 MHz by a pulse picker system. After passing thought an immersion-oil objective with the numerical aperture of ~1.3, the laser excitation beam was focused onto the sample substrate with a spot size around ~1 μm. The optical emission from a single FAPbI$_3$/NdF$_3$ NC was collected by the same objective and sent sequentially to a spectrometer and a charge-coupled device (CCD) for the PL spectral measurement with an integration time of 1 s. Alternatively, after passing through a 50/50 non-polarizing beam splitter, the optical emission from a single FAPbI$_3$/NdF$_3$ NC could be sent to two avalanche photodetectors (APDs) to obtain the PL intensity time trace and the PL decay curve with a time resolution of ~100 ps.

**TA measurement.** One drop of the concentrated sample solution was spin-coated onto a SiO$_2$ substrate to form a solid film of the ensemble FAPbI$_3$/NdF$_3$ NCs. The ~800 nm output from an amplified Ti:Sapphire laser, with the repetition rate of ~1 KHz and the pulse width of ~100 fs, was split into two beams by a non-polarizing beam splitter. One beam was sent to an optical parametric amplifier (OPA) to get the pump at either ~640 or ~355 nm, while the other beam was focused onto a sapphire plate to generate the probe of a white-light continuum. The pump beam was first sent to the optical line capable of changing the pulse time delay from ~200 fs to ~1 ns, and then focused by a SiO$_2$ lens (focal length, ~10 cm) onto the FAPbI$_3$/NdF$_3$ NCs with a spot size of ~1.4 mm and an incident angle of ~45° relative to the surface normal. The probe beam was reflected by a parabolic mirror (focal length, ~7.5 cm) to overlap with the pump beam with a relatively smaller spot size ~0.5 mm and an incident angle of ~0° relative to the surface normal. After passing through the FAPbI$_3$/NdF$_3$ NCs, the transmitted probe beam was dispersed by a grating and then arrived at a linear-array CMOS detector. Since the pump beam was modulated at ~500 Hz by a mechanical optical chopper right after leaving OPA, the probe beam spectra could be obtained with and without its presence to yield a TA spectrum for the transmission change at any given time delay.

**ASE measurement.** As in the TA measurement, one drop of the concentrated sample solution was spin-coated onto a SiO$_2$ substrate to form a solid film of the ensemble FAPbI$_3$/NdF$_3$ NCs. The ~800 nm output beam from an amplified Ti:Sapphire laser, with a repetition rate of ~1 KHz and a pulse width of ~100 fs, was sent to an OPA to get the excitation wavelength at either ~640 or ~355 nm. After passing through a SiO$_2$ lens with the focal length of ~10 cm, the ~640 (~355) nm laser beam was focused vertically onto the sample substrate with the spot size of ~1.2 (~1.0) mm. The optical emission from the ensemble FAPbI$_3$/NdF$_3$ NCs was collected from the solid film edge by a dry objective with a numerical aperture of ~0.55, and then sent to a spectrometer and a linear-array CCD for the PL and ASE spectral measurements with an integration time of 1 s.

**Data availability**

The data supporting the findings of this study are available from the corresponding authors upon request.

**Acknowledgements**

This work is supported by the National Basic Research Program of China (2021YFA1400803), and the National Natural Science Foundation of China (12574456, 12274216 and 62174081).

**Author contributions**

F.H., M.X. and X.W. conceived and designed the experiments. Z.Z., E.S. and J.L. prepared the samples and performed the optical measurements. Z.Z., C.Z., F.H. and X.W. contributed to the discussions. Z.Z. F.H. and X.W. co-wrote the manuscript.

**Competing Interests**

The authors declare no competing financial or non-financial interests.

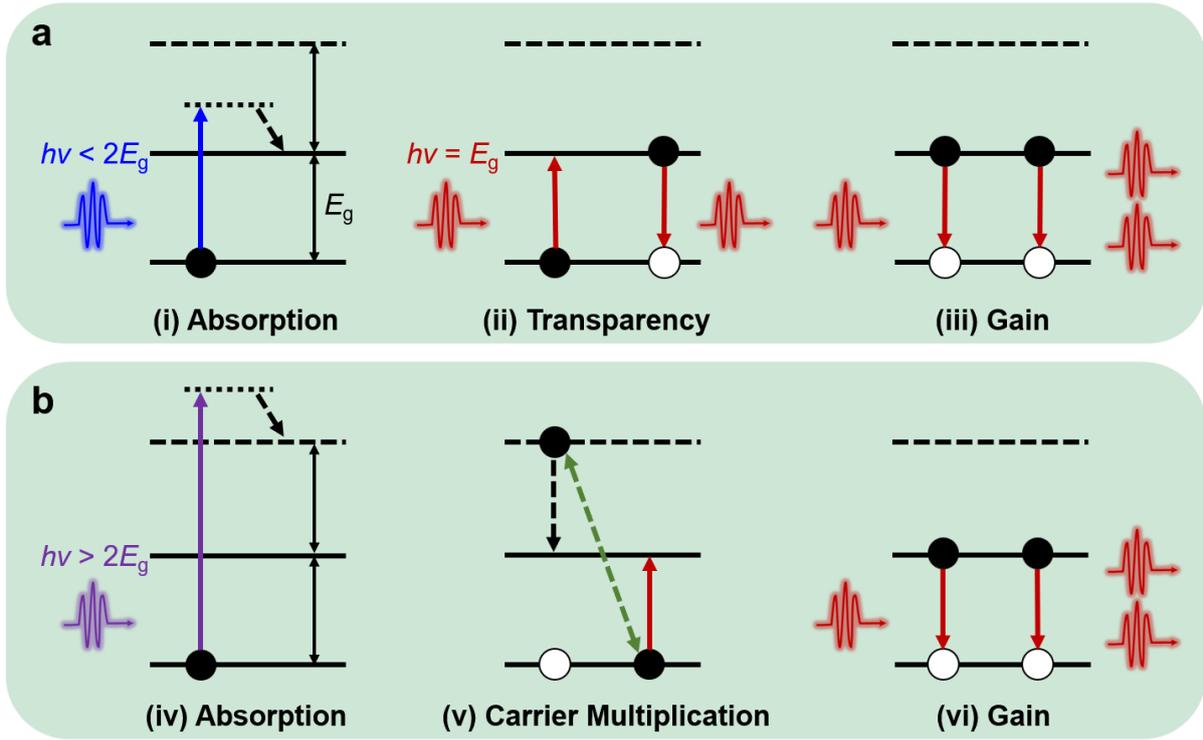

**Fig. 1 | Optical gain in a single NC. a,** (i) Under low-power laser excitation ($hv < 2\,E_g$) of a single NC, an electron is promoted from the valence band to a lower-lying excited state in the conduction band and then relaxes to the band-edge state to finish the optical absorption process. (ii) This single NC is optically transparent to the incident photon ($hv = E_g$), since the probabilities for it to induce SE from the band-edge electron and to excite another valence-band electron are equal to each other. (iii) Under high-power laser excitation ($hv < 2\,E_g$) of a single NC to create two band-edge electrons, the optical gain is realized so that one incident photon can induce two emitted photons all with the same energy of $hv = E_g$. **b,** (iv) Under low-power laser excitation ($hv > 2\,E_g$) of a single NC, an electron is promoted from the valence band to a higher-lying excited state in the conduction band to finish the optical absorption process. (v) This electron arrives directly at the band-edge state after a slight relaxation, with the released energy exciting another valence-band electron in the CM process. (vi) The band-edge state is now occupied by two electrons to realize optical gain in a single NC, upon the absorption of only a single laser photon with the energy of $hv > 2\,E_g$.

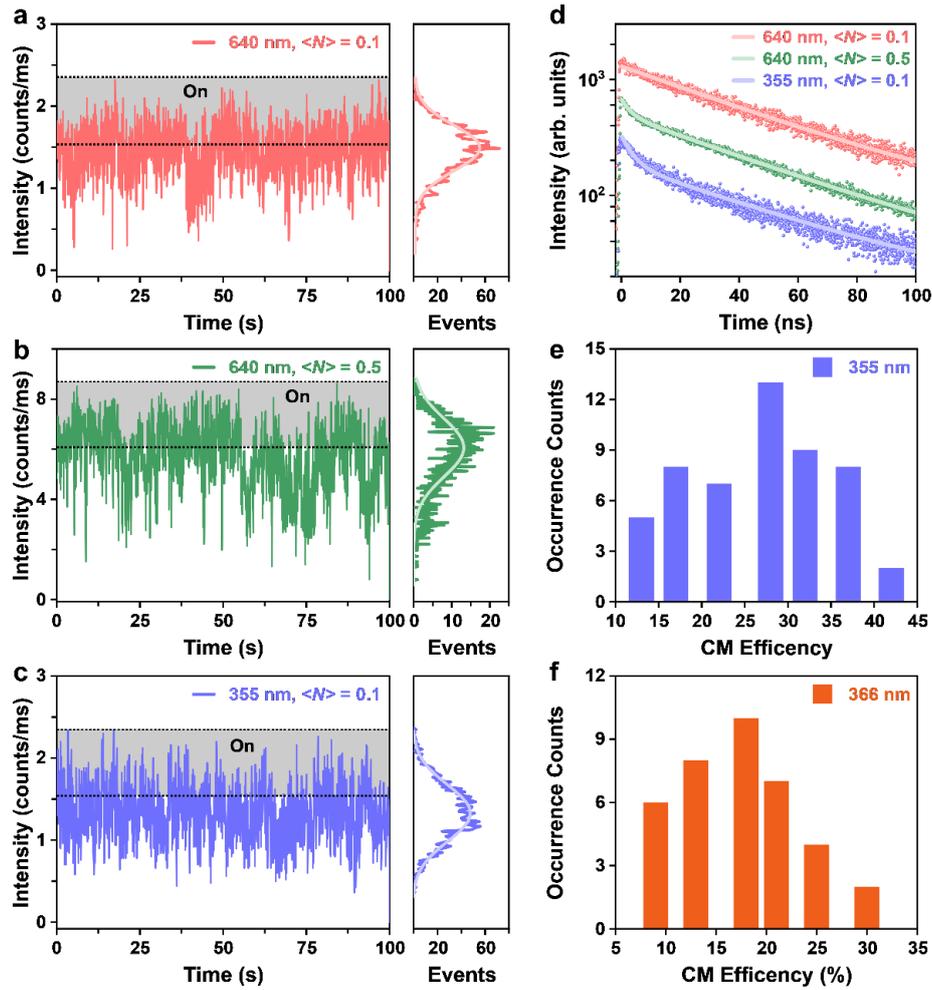

**Fig. 2 | CM in the single FAPbI$_3$/NdF$_3$ NCs.** PL intensity time traces of a single FAPbI$_3$/NdF$_3$ NC excited at **a,** ~640 nm with $<N> = $ ~0.1, **b,** ~640 nm with $<N> = $ ~0.5 and **c,** ~355 nm with $<N> = $ ~0.1, with the corresponding single-peak Gaussian distributions being shown in the right panels. **d,** PL decay curves extracted from **a-c** for the neutral-exciton photons as marked by the grey "on" boxes, and fitted by either a single-exponential function (~640 nm, $<N> = $ ~0.1) or a biexponential function (~640 nm, $<N> = $ ~0.5 and ~355 nm, $<N> = $ ~0.1). **e,** Statistical histogram for the distribution of CM efficiencies measured for 52 single FAPbI$_3$/NdF$_3$ NCs (~355 nm, $<N> = $ ~0.1) with an average value of ~25.7%. **f,** Statistical histogram for the distribution of CM efficiencies measured for 37 single FAPbI$_3$/NdF$_3$ NCs (~366 nm, $<N> = $ ~0.1) with an average value of ~18.1%.

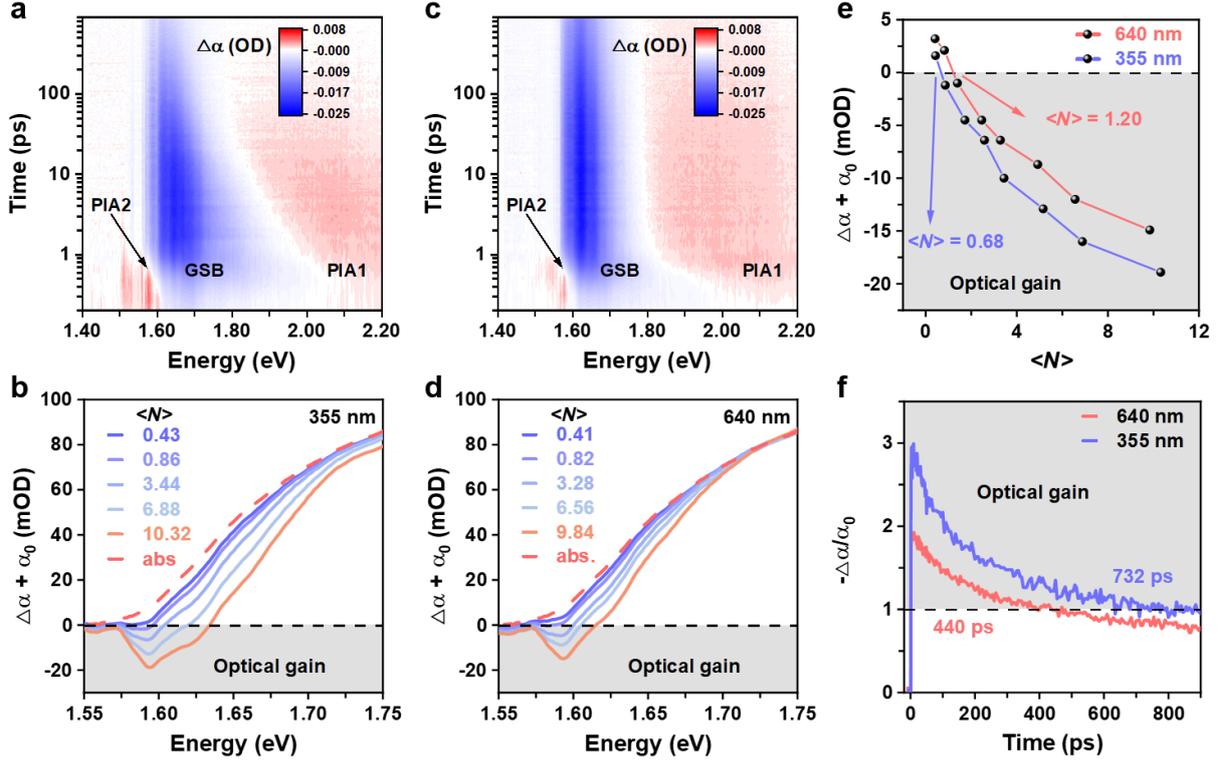

**Fig. 3 | Optical gain of the ensemble FAPbI$_3$/NdF$_3$ NCs.** 2D images showing time-dependent evolutions of the TA spectra measured under the **a,** ~355 nm and **c,** ~640 nm laser excitations with $<N>$ = ~10.32 and ~9.84, respectively. Nonlinear absorption spectra of $\Delta\alpha + \alpha_0$ obtained at the time delay of ~4 ps under the **b,** ~355 nm and **d,** ~640 nm laser excitations with increasing $<N>$ values, together with the steady-state absorption spectra (abs.). **e,** $<N>$-dependent $\Delta\alpha + \alpha_0$ values obtained at the time delay of ~4 ps for the ~1.59 eV biexciton peak, from which the optical gain thresholds of $<N>$ = ~0.68 and ~1.20 can be estimated under the ~355 and ~640 nm laser excitations, respectively. **f,** Time-dependent $-\Delta\alpha/\alpha_0$ curves for the ~1.59 eV biexciton peak, from which the optical gain lifetimes of ~732 and ~440 ps can be estimated under the ~355 and ~640 nm laser excitations with $<N>$ = ~10.32 and ~9.84, respectively.

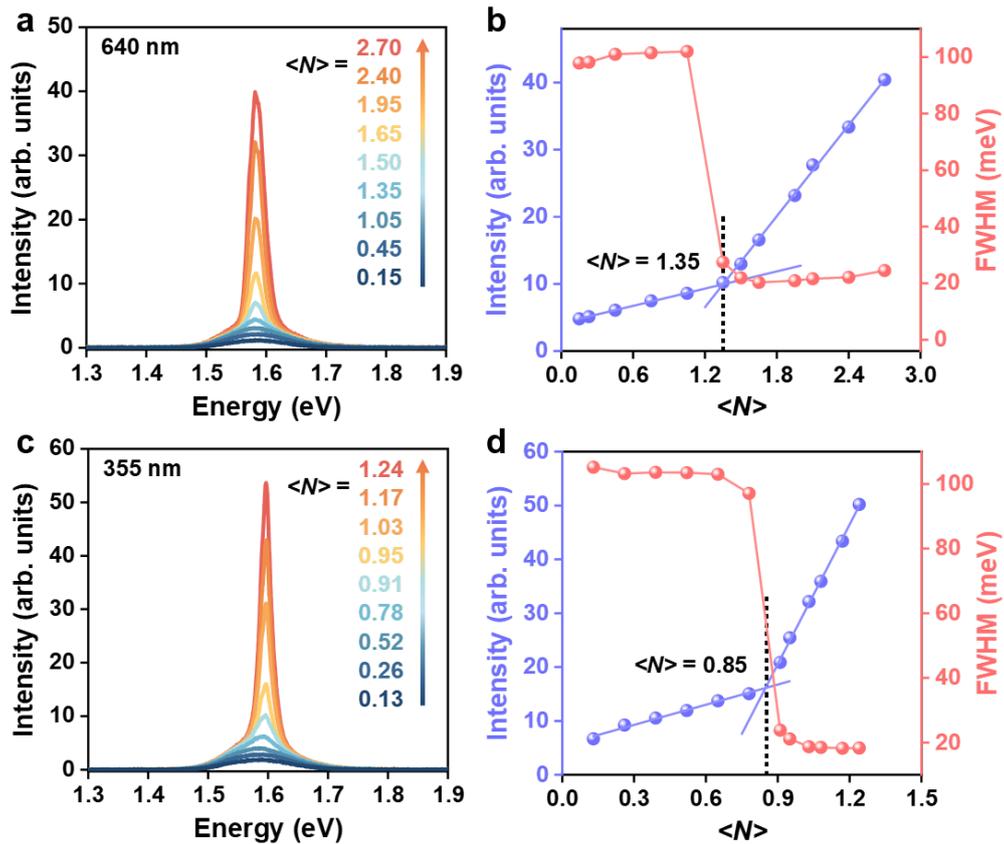

**Fig. 4 | ASE of the ensemble FAPbI₃/NdF₃ NCs. a**, PL and ASE spectra measured for the FAPbI₃/NdF₃ NCs under the ~640 nm laser excitation with increasing <N> values from ~0.15 to ~2.70. **b,** <N>-dependent spectral integrated intensities and peak FWHMs extracted from **a**, from which an ASE threshold of <N> = ~1.35 can be determined for the ~640 nm laser excitation. **c,** PL and ASE spectra measured for the FAPbI₃/NdF₃ NCs under the ~355 nm laser excitation with increasing <N> values from ~0.13 to ~1.24. **d,** <N>-dependent spectral integrated intensities and peak FWHMs extracted from **c**, from which an ASE threshold of <N> = ~0.85 can be determined for the ~355 nm laser excitation.

**Supplementary Information**

**Reduced Optical Gain Threshold by Carrier Multiplication in Semiconductor Perovskite Nanocrystals**

Zhen Zhang[1†], Encheng Sun[1†], Jian Li[1†], Chunfeng Zhang[1], Fengrui Hu[2*], Min Xiao[1], and Xiaoyong Wang[1*]

[1]*National Laboratory of Solid State Microstructures, School of Physics, and Collaborative Innovation Center of Advanced Microstructures, Nanjing University, Nanjing 210093, China*

[2]*College of Engineering and Applied Sciences, and MOE Key Laboratory of Intelligent Optical Sensing and Manipulation, Nanjing University, Nanjing 210093, China*

[*]Correspondence to F.H. (frhu@nju.edu.cn) or X.W. (wxiaoyong@nju.edu.cn)

[†]These authors contributed equally to this work

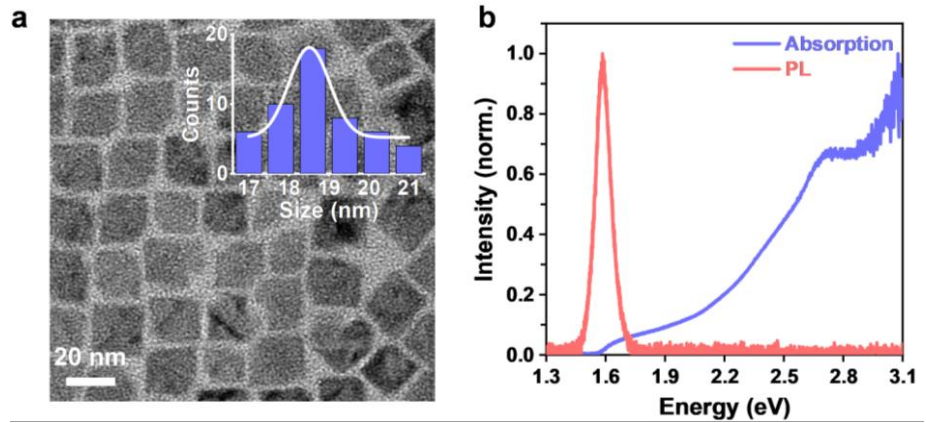

**Supplementary Fig. 1. a,** Transmission electron microscope image measured for the FAPbI$_3$/NdF$_3$ NCs. Inset: Statistical distribution for the edge lengths of the FAPbI$_3$/NdF$_3$ NCs with an average value of ~18.5 nm. **b,** Solution absorption and PL spectra measured for the FAPbI$_3$/NdF$_3$ NCs with the band-edge peaks located at ~1.62 and ~1.58 eV, respectively.

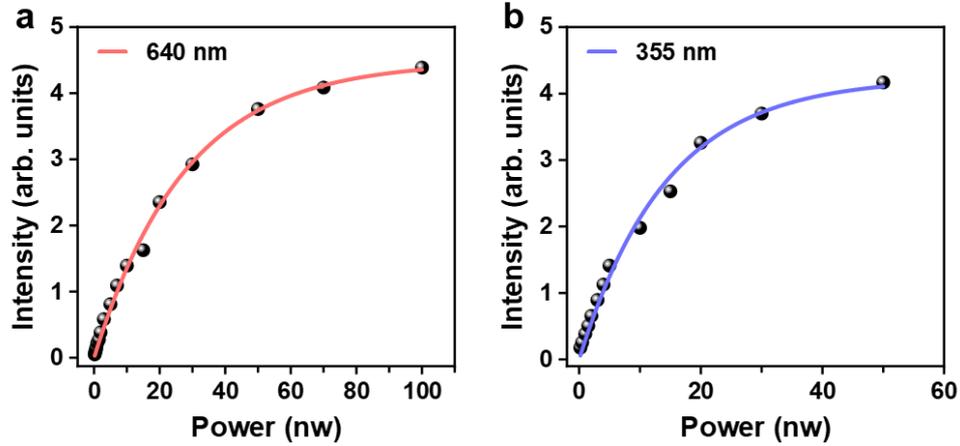

**Supplementary Fig. 2. a,** PL intensity ($I$) of the ensemble FAPbI$_3$/NdF$_3$ NCs measured as a function of the ~640 nm pulsed laser power ($P$) and fitted with the function of $I \propto 1 - e^{-<N>} = 1 - e^{-\alpha P}$. Here, $<N>$ is the average number of photons absorbed per pulse by a single NC, while $\alpha$ is a fitting constant related to the absorption cross section. After getting $\alpha$ = ~0.036 nW$^{-1}$, the $<N>$ value at any given $P$ of the ~640 nm excitation laser can be calculated. **b,** PL intensity ($I$) of the ensemble FAPbI$_3$/NdF$_3$ NCs measured as a function of the ~355 nm pulsed laser power ($P$) and fitted with the function of $I \propto 1 - e^{-<N>} = 1 - e^{-\alpha P}$. After getting $\alpha$ = ~0.070 nW$^{-1}$, the $<N>$ value at any given $P$ of the ~355 nm excitation laser can be calculated. Similar laser power-dependent PL intensities have been measured for the ensemble FAPbI$_3$/NdF$_3$ NCs at other laser excitation wavelengths (such as ~366 and ~385 nm) for the estimation of $<N>$ at a given $P$.

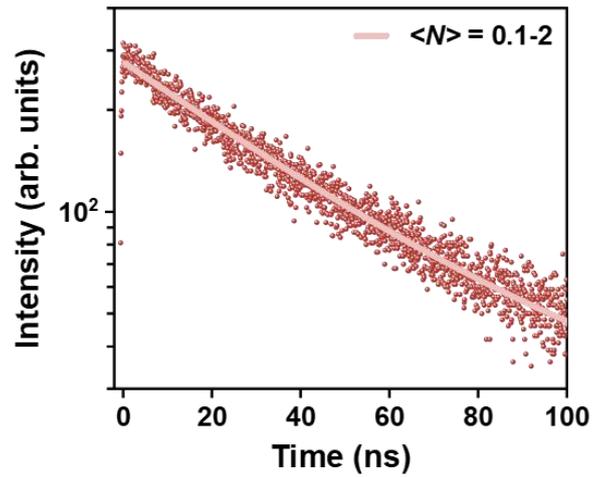

**Supplementary Fig. 3.** PL decay curve measured for a single FAPbI$_3$/NdF$_3$ NC under the ~640 nm laser excitation with $<N>$ = ~0.1, and fitted by a single-exponential function with the lifetime of ~45.7 ns. Before the above measurement, this single FAPbI$_3$/NdF$_3$ NC has been excited at ~355 nm with $<N>$ = ~0.1 for the PL decay curve measurement.

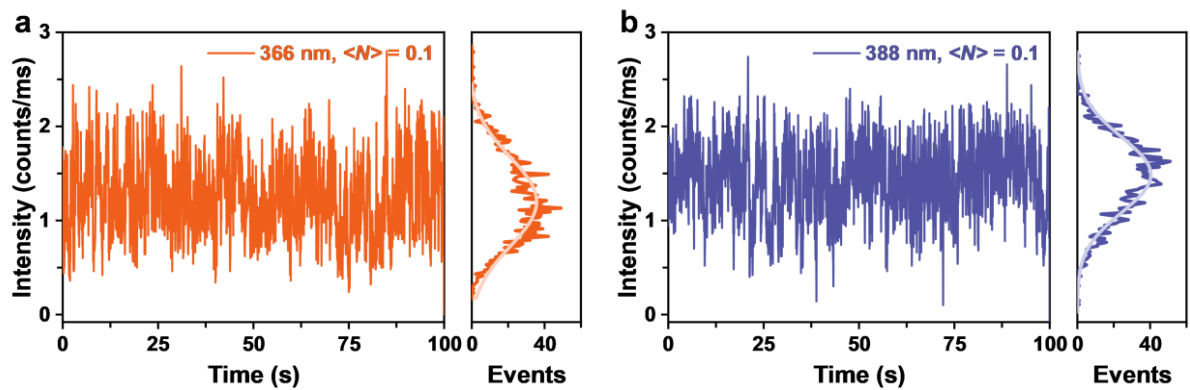

**Supplementary Fig. 4.** Time-dependent PL intensities of a single FAPbI$_3$/NdF$_3$ NC excited at **a,** ~366 nm and **b,** ~388 nm with $<N>$ = ~0.1, with the corresponding single-peak Gaussian distributions being shown in the right panels.

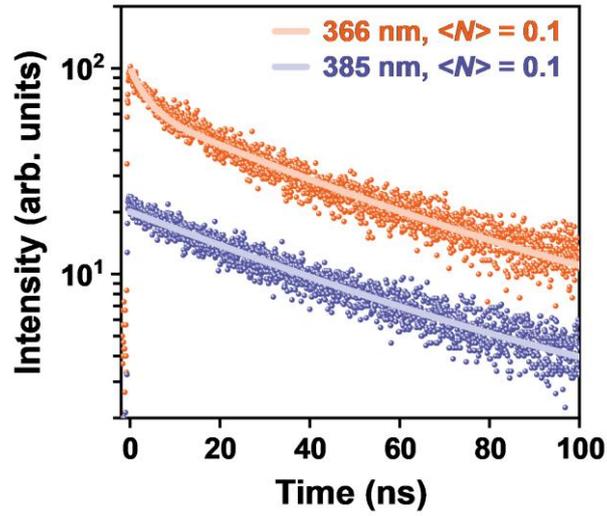

**Supplementary Fig. 5.** PL decay curves extracted from Supplementary Fig. 4 for the neutral-exciton photons of a single $FAPbI_3/NdF_3$ NC excited with $<N>$ = ~0.1 at the ~366 and ~388 nm laser wavelengths, respectively. The PL decay curve obtained at ~366 nm is fitted by a biexponential function with the slow and fast lifetimes of ~46.6 and ~3.7 ns, respectively, while that obtained at ~388 nm is fitted with a single-exponential lifetime of ~46.2 ns.

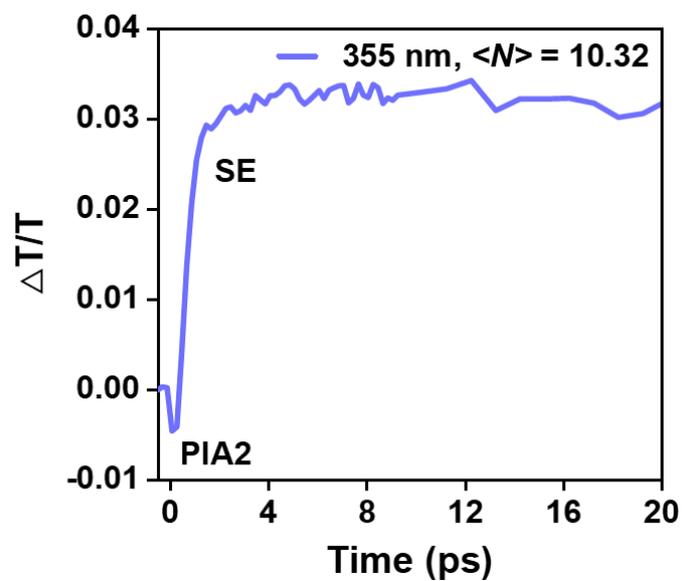

**Supplementary Fig. 6.** TA curve measured at the biexciton peak of ~1.59 eV for the ensemble FAPbI$_3$/NdF$_3$ NCs under the ~355 nm laser excitation with $<N>$ = ~10.32, showing the transition from PIA2 to SE and the arrival at a highest level for the latter signal at the time delays of ~1 and ~4 ps, respectively.